\documentclass[useAMS,usenatbib]{mn2e}
\usepackage[utf8]{inputenc}
\usepackage[T1]{fontenc}
\usepackage{lmodern}
\usepackage{graphicx}
\usepackage{subfigure}
\usepackage{natbib}
\usepackage{hyperref}

\newcommand{\lum}{{\rm erg\,s^{-1}}}
\newcommand{\nar}{NewAR}
\newcommand{\actaa}{AcA}
\newcommand{\aap}{A\&A}
\newcommand{\mnras}{MNRAS}
\newcommand{\apj}{ApJ}
\newcommand{\apjs}{ApJS}
\newcommand{\pasp}{PASP}
\newcommand{\aj}{AJ}
\newcommand{\iaucirc}{IAU Circ}
\newcommand{\apjl}{ApJL}
\newcommand{\nat}{Nature}

\begin{document}

\title[Duty Cycle, X-ray luminosity, and Specific Frequency of DNe]{The Relationship Between X-ray Luminosity and Duty Cycle for Dwarf Novae and their Specific Frequency in the Inner Galaxy}
\date{2014 September 08}
\pubyear{0000} \volume{000} \pagerange{1}

\author[C.~T. Britt et al.]{C.~T.~Britt,$^{1}$
T.~Maccarone,$^{1}$
M.~L.~Pretorius,$^{2}$
R.~I.~Hynes,$^{3}$
P.~G.~Jonker,$^{4,5,6}$
\newauthor
M.~A.~P.~Torres,$^{4,5}$
C.~Knigge,$^{7}$
C.O. Heinke$^{8,9}$
C.~B.~Johnson,$^{3}$
D.~Steeghs,$^{10}$
\newauthor
S.~Greiss,$^{10}$
G.~Nelemans,$^{5}$\\
$^{1}$Texas Tech University, Department of Physics, Box 41051, Science Building, Lubbock, TX 79409-1051, USA\\
$^{2}$Oxford University, Department of Physics, Denys Wilkinson Building, Keble Road, Oxford, OX1 3RH, UK\\
$^{3}$Louisiana State University, Department of Physics and Astronomy, Baton Rouge LA 70803-4001, USA\\
$^{4}$SRON, Netherlands Institute for Space Research, Sorbonnelaan 2, 3584~CA, Utrecht, The Netherlands\\
$^{5}$Dept. of Astrophysics, IMAPP, Radboud University Nijmegen, Heyendaalseweg 135, 6525 AJ, Nijmegen, The Netherlands\\
$^{6}$Harvard-Smithsonian Center for Astrophysics, 60 Garden Street, Cambridge, MA~02138, U.S.A.\\
$^{7}$School of Physics and Astronomy, University of Southampton, Hampshire SO17 1BJ, United Kingdom \\
$^8$University of Alberta, Physics Dept., CCIS 4-183, Edmonton, AB, T6G 2E1, Canada \\
$^{9}$Max-Planck-Institute f{\"u}r Radioastronomie, Auf dem H{\"u}gel 69, 53121, Bonn, Germany \\
$^{10}$Astronomy and Astrophysics, Dept. of Physics, University of Warwick, Coventry, CV4~7AL, United Kingdom}

\maketitle

\begin{abstract}

We measure the duty cycles for an existing sample of well observed, nearby dwarf novae using data from AAVSO, and present a quantitative empirical relation between the duty cycle of dwarf novae outbursts and the X-ray luminosity of the system in quiescence. We have found that $\log DC=0.63(\pm0.21)\times(\log L_{X}(\lum)-31.3)-0.95(\pm0.1)$, where DC stands for duty cycle. We note that there is intrinsic scatter in this relation greater than what is expected from purely statistical errors. Using the dwarf nova X-ray luminosity functions from \citet{Pretorius12} and \citet{Byckling10}, we compare this relation to the number of dwarf novae in the Galactic Bulge Survey which were identified through optical outbursts during an 8-day long monitoring campaign.  We find a specific frequency of X-ray bright ($L_{X}\ga10^{31}\,\lum$) Cataclysmic Variables undergoing Dwarf Novae outbursts in the direction of the Galactic Bulge of $6.6\pm4.7\times10^{-5}\,M_{\odot}^{-1}$. Such a specific frequency would give a Solar neighborhood space density of long period CVs of $\rho=5.6\pm3.9\times10^{-6}\,$pc$^{-3}$. We advocate the use of specific frequency in future work, given that projects like LSST will detect DNe well outside the distance range over which $\rho\approx{\textrm const}$.

\end{abstract}


\vspace{-3em}
\section{Introduction}
\label{intro}

Cataclysmic variables (CVs) are binary star systems in which a white dwarf (WD) 
primary accretes material from a low mass secondary star through Roche-lobe 
overflow. In the absence of strong magnetic fields of the WD, the infalling 
material forms an accretion disk, through which angular momentum can be 
transfered between particles allowing accretion onto the surface of the WD. 

The material in the disk heats as it builds up, eventually reaching the 
ionization temperature of hydrogen. At this point, the gas in the disk becomes 
much more viscous \citep{Smak99,Lasota01}. The heat energy from the increase 
in viscosity causes a heating wave to propagate through the disk, ionizing the 
remainder of it. Mass then passes through the high viscosity disk at a much 
higher rate, dumping material onto the WD. A sudden brightening of 
several magnitudes and subsequent decline to the original brightness is 
observed over the course of days to weeks \citep{Smak71,Warner71}. These Dwarf 
Novae (DNe) occur with a 
frequency tied to the recent mass accretion rate $\dot{M_{1}}$ and mass transfer
rate $\dot{M_{2}}$ of the system, as the disk must refill in between outbursts.
High mass transfer rate CVs (nova-like objects) do not undergo DNe as the disks are 
constantly above the ionization temperature of hydrogen. Low mass transfer rate 
objects such as WZ Sge may go decades between recurrences of DN outbursts. 

The mass accretion rate in CVs is also known to be tied to X-ray luminosity up 
to a critical rate at which point the Boundary Layer (BL) on the 
surface of the WD becomes opaque to X-rays \citep{Patterson85,Wheatley00,Baskill01}. Indeed, the X-ray 
emission during a DN outburst can rise at the start of the outburst as for SS Cyg 
until the BL becomes opaque at which point the X-rays are quenched. 
Some DNe never reach this critical accretion rate and the 
X-ray emission remains unquenched through the duration of the outburst 
\citep[e.g. U Gem;][]{Swank78}.

The empirical population of CVs can constrain evolution models; a crucial measurement to make with this end in mind is the space density of CVs, $\rho$. Models of binary evolution predict space densities ranging over an order of magnitude or more, from $10^{-5}-10^{-4}$\,pc$^{-3}$ \citep{deKool92,Kolb93,Politano96}. Observational campaigns, largely using X-ray selected samples, have made estimates of the CV space density converging on 0.5-1$\times10^{-5}$ pc$^{-3}$ \citep{Patterson98,Schreiber03,Aungwerojwit06,Pretorius07,Rogel08,Pretorius12}.  
The space density found in \citet{Pretorius12} for long-period CVs was $2.1^{+3.5}_{-1.3}\times10^{-6}$ pc$^{-3}$ which assumed that all systems with unknown periods had periods below the period gap. Studies of the space density of CVs have focused on the solar neighborhood, for the simple and expedient reason that a complete census of faint CVs will be impossible to acquire at large distances. The outlier among estimates of space density from observational campaigns, finding a much smaller space density for the dwarf nova population ($<5\times10^{-7}$ pc$^{-3}$), used the OGLE survey towards the Galactic Bulge to find dwarf nova outbursts \citep{Cieslinski03}. This estimate was based on identifying distant DN outbursts as part of a larger wide-field survey as is this work, where \citet{Cieslinski03} uses the OGLE-II survey and this work is based on the Chandra Galactic Bulge Survey (GBS).

The GBS is an X-ray survey between $-3^{\circ}<l<3^{\circ}$ and $1^{\circ}<|b|<2^{\circ}$ with shallow coverage of 2\,ks \citep{Jonker11,Jonker14}. The primary goal of the GBS is to conduct a census of Low Mass X-ray Binaries in quiescence (qLMXB), in order to constrain models of binary evolution and to identify new systems for which mass determinations are possible, including eclipsing systems. Towards the goal of classifying each X-ray source in the GBS, multiwavelength photometric and spectroscopic followup campaigns have been undertaken. Some of these focus on classifying new interesting sources with detailed followup, such as new CVs \citep{Britt13,Ratti13}, a possible new black hole binary \citep{Britt13}, a Slowly Pulsating B star with a white dwarf or coronally active G-type companion \citep{Johnson14}, and a likely Carbon star Symbiotic Binary \citep{Hynes14}. In terms of more comprehensive followup, \citet{Britt14} focuses on variable optical counterparts to GBS sources including the identification of DNe in outburst; \citet{Greiss13} identifies the infrared counterparts; \citet{Hynes12} cross matches GBS sources with bright star catalogs such as Tycho and ASAS; \citet{Maccarone12} cross matches GBS sources with existing catalogs of radio sources; \citet{Torres14} and \citet{Wu14} provide optical spectroscopy of stars with an H$\alpha$ excess near GBS sources. The initial X-ray survey, which covered 3/4 of the planned survey area, discovered 1216 unique X-ray sources with 3 or more detected photons in a 2\,ks observation. The X-ray observations were kept intentionally shallow in order to keep the number of active stars and CVs in the survey low while still reaching the Bulge for neutron star and long period black hole qLMXBs. As a result, the non-magnetic CVs in the GBS are generally closer than Bulge distance. For example, the X-ray bright system Z Cam would be detected up to $\sim4500$\,pc away.

Our use of an optical variability survey as part of the GBS to constrain the space density of CVs in the direction of the Galactic Bulge mirrors in some ways that undertaken with the OGLE-II catalog in \citet{Cieslinski03}. Our survey differs in several key respects. First, our analysis takes into account the likelihood of detecting dwarf novae with recurrence times much longer than the baseline of observations. Not allowing for dwarf novae with recurrence periods longer than the OGLE time baseline is the likely reason why \citet{Cieslinski03} obtained a final DN space density much lower than that obtained by other methods, or that we find in this paper. Second, the limiting magnitude in our survey is slightly deeper than that of OGLE-II, $r'=23$ versus $I=22$, though this is equalized by reddening. Third, our baseline of observations is dramatically shorter than in OGLE-II, with an 8 day baseline compared to OGLE-II's 4 years of observations, though both are shorter than the recurrence time of many CVs undergoing DN outbursts and are therefore far from complete samples. Fourth, our optically surveyed area is currently somewhat smaller, 9 square degrees \citep{Britt14} instead of OGLE-II's 12. Lastly, our optical detections of DNe, at the present time, are premised on an X-ray detection of the same CV. Recent progress in determining the X-ray luminosity function of CVs combined with a more complete sample of well observed, nearby DNe allows our sample to be leveraged in ways that were not available at the time of OGLE-II as well. \citet{Cieslinski03} place an upper limit on the space density of CVs towards the Bulge of $5\times10^{-7}$\,pc$^{-3}$. In this work, we reconcile recent estimates of the X-ray selected CV space density with an optical-outburst selected dwarf nova space density.

\vspace{-2em}
\section{Duty Cycle vs X-ray Luminosity}

\subsection{Sample Selection and Measurements for Duty Cycle versus Luminosity Relation}

As a basis for determining the relationship between X-ray luminosities of CVs and other key properties, we use the survey of DN within $\sim200$\,pc in \citet{Byckling10}. This sample, although small (13 CVs), includes only objects with accurate distance measurements from trigonometric parallax. \citet{Byckling10} demonstrate that this sample is likely suffering from selection effects against low luminosities ($L_{X}<3\times10^{30}\,\lum$ in the $[2-10]$\,keV band) and duty cycles. We therefore limit our fit to the CVs with an X-ray luminosity above this cutoff. This sample is not distance complete, as it is limited to sources with parallax measurements which were mostly done by the same group with Northern hemisphere telescopes \citep{Thorstensen03,Thorstensen08,Harrison04}. As an exercise, we have fit the entire sample as well and find the resulting relation to be consistent with the more narrowly defined sample.

Since the publication of \citet{Byckling10}, the distance to SS Cyg has been revised downward by the improved parallax measure in \citet{Millerjones13}, from $159\pm12$\,pc to $114\pm2$\,pc. This results in a drop in the X-ray luminosity by a factor of about 2 compared to the value listed in \citet{Byckling10}. 

\begin{figure*}[p!]
\begin{minipage}{0.9\textwidth}
\centering
\parbox{\textwidth}{
\subfigure{\includegraphics[width=0.25\textwidth,angle=90]{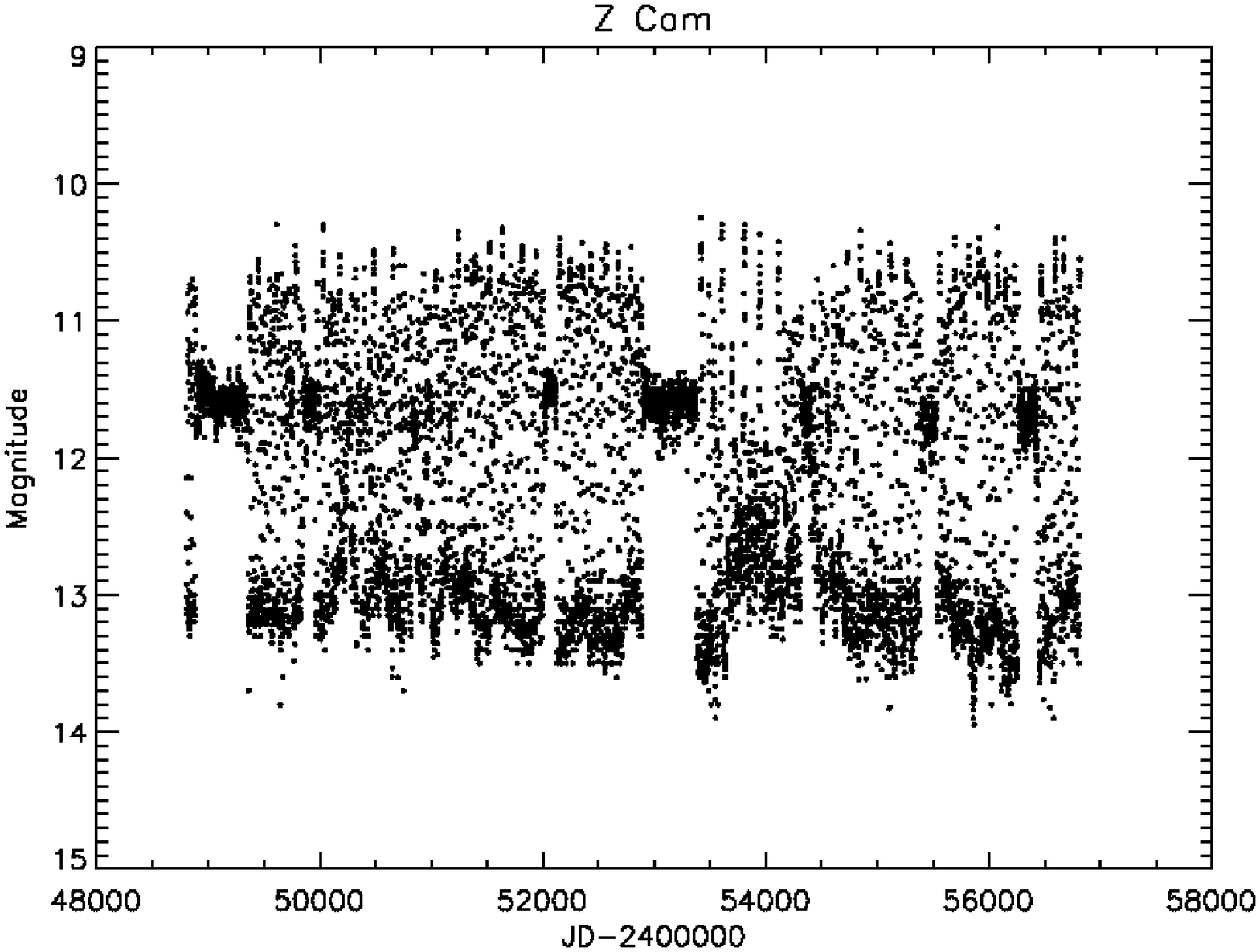}}
\subfigure{\includegraphics[width=0.25\textwidth,angle=90]{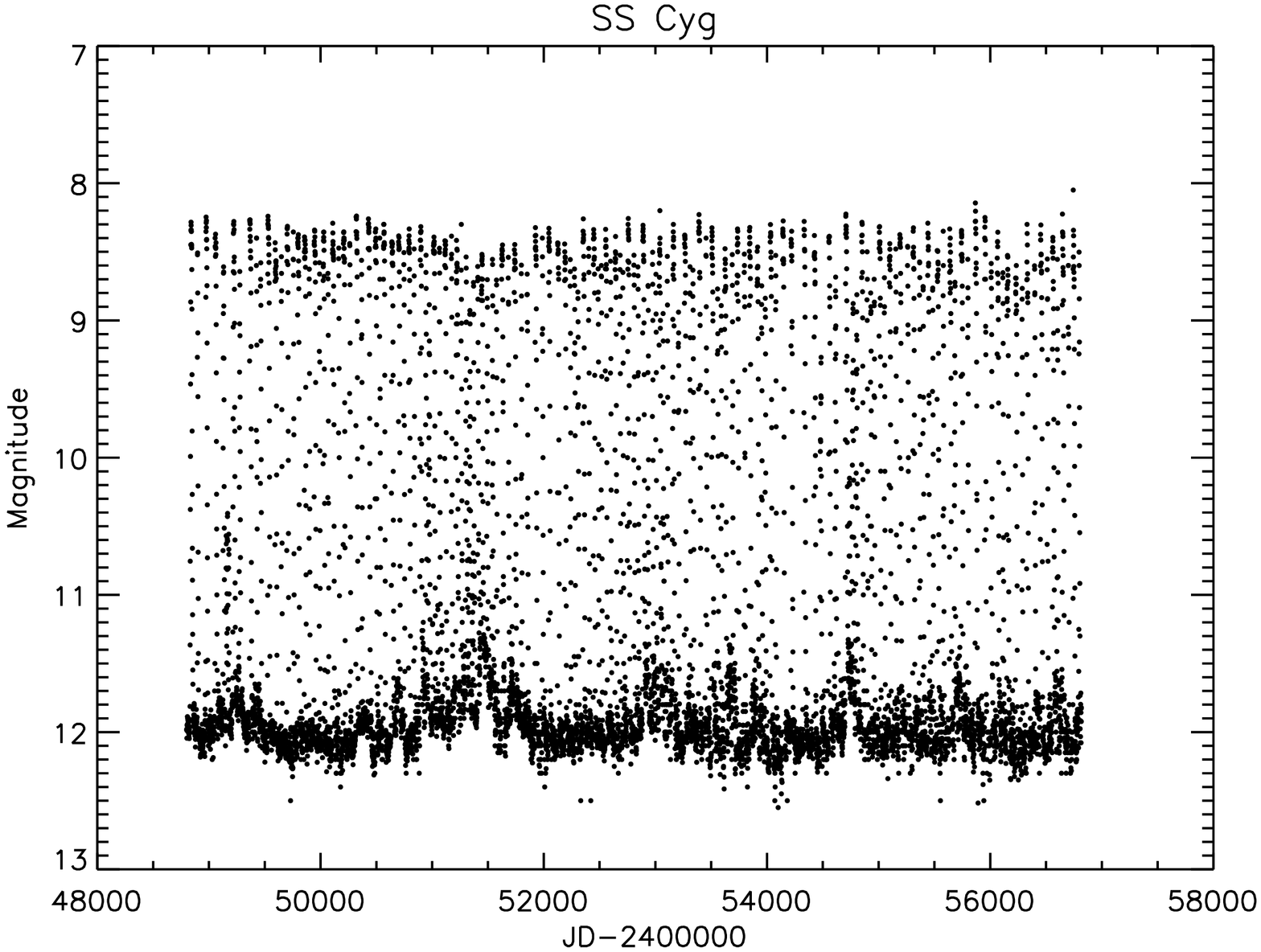}}
\subfigure{\includegraphics[width=0.25\textwidth,angle=90]{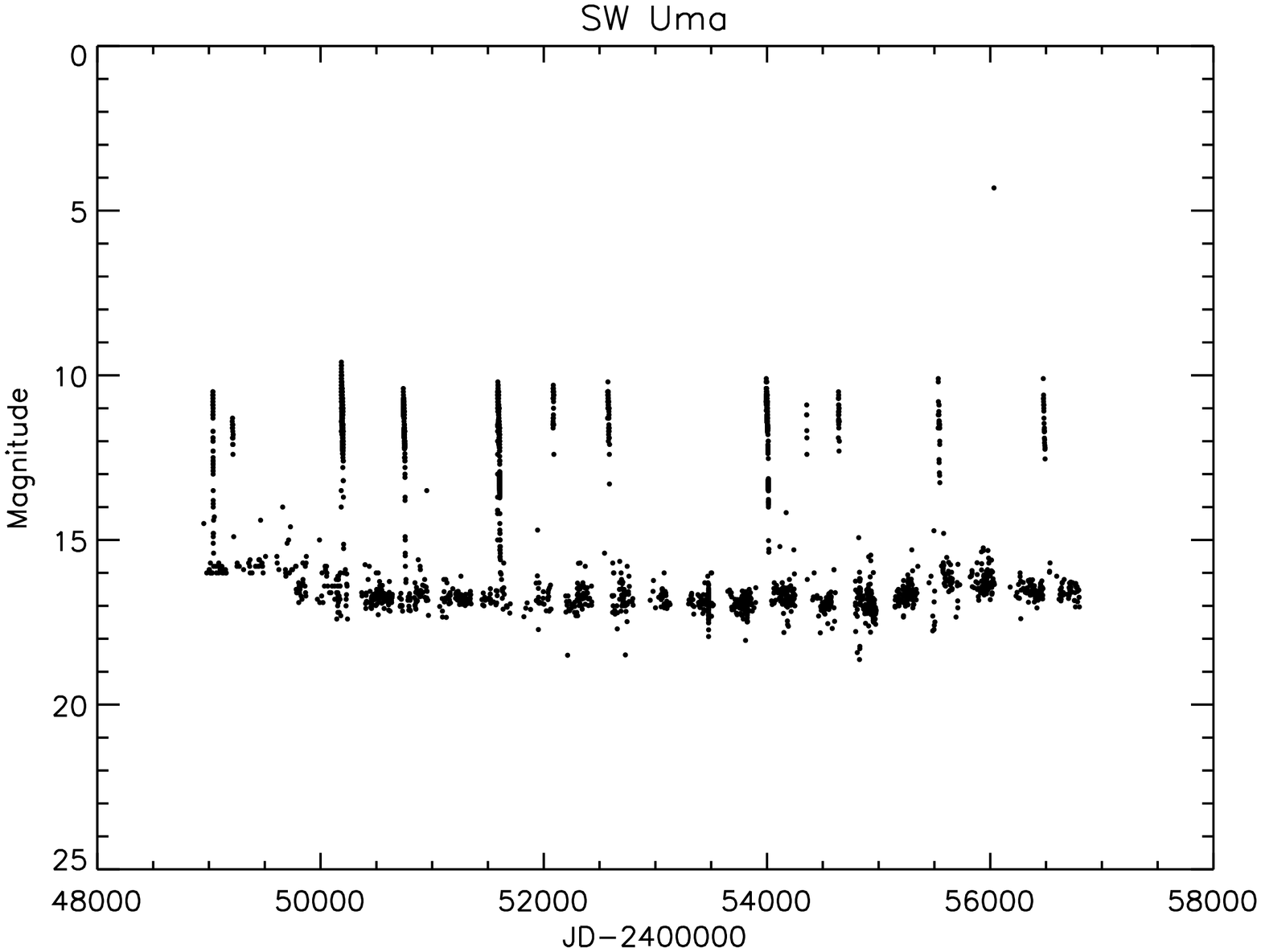}} \\
\subfigure{\includegraphics[width=0.25\textwidth,angle=90]{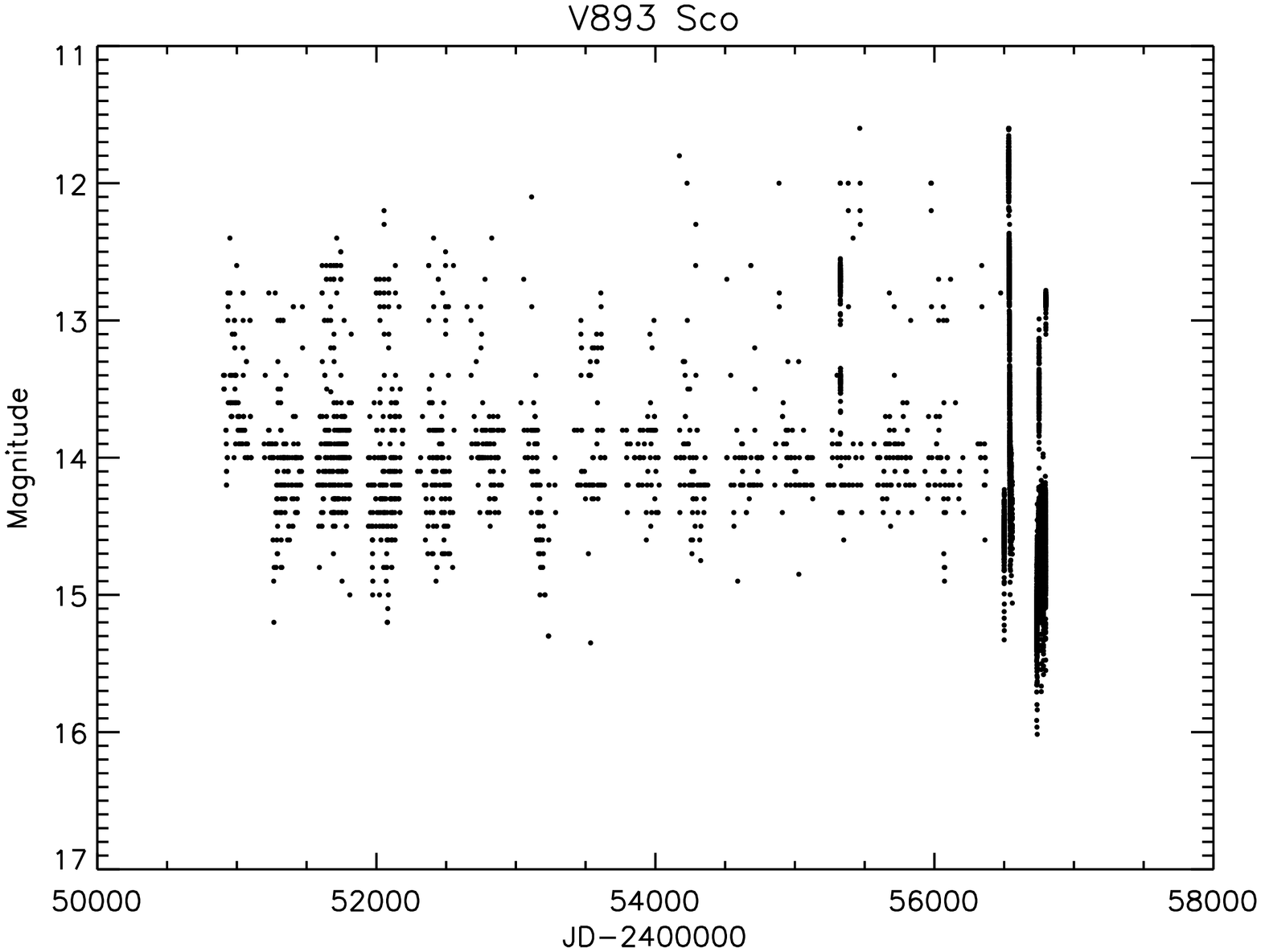}}
\subfigure{\includegraphics[width=0.25\textwidth,angle=90]{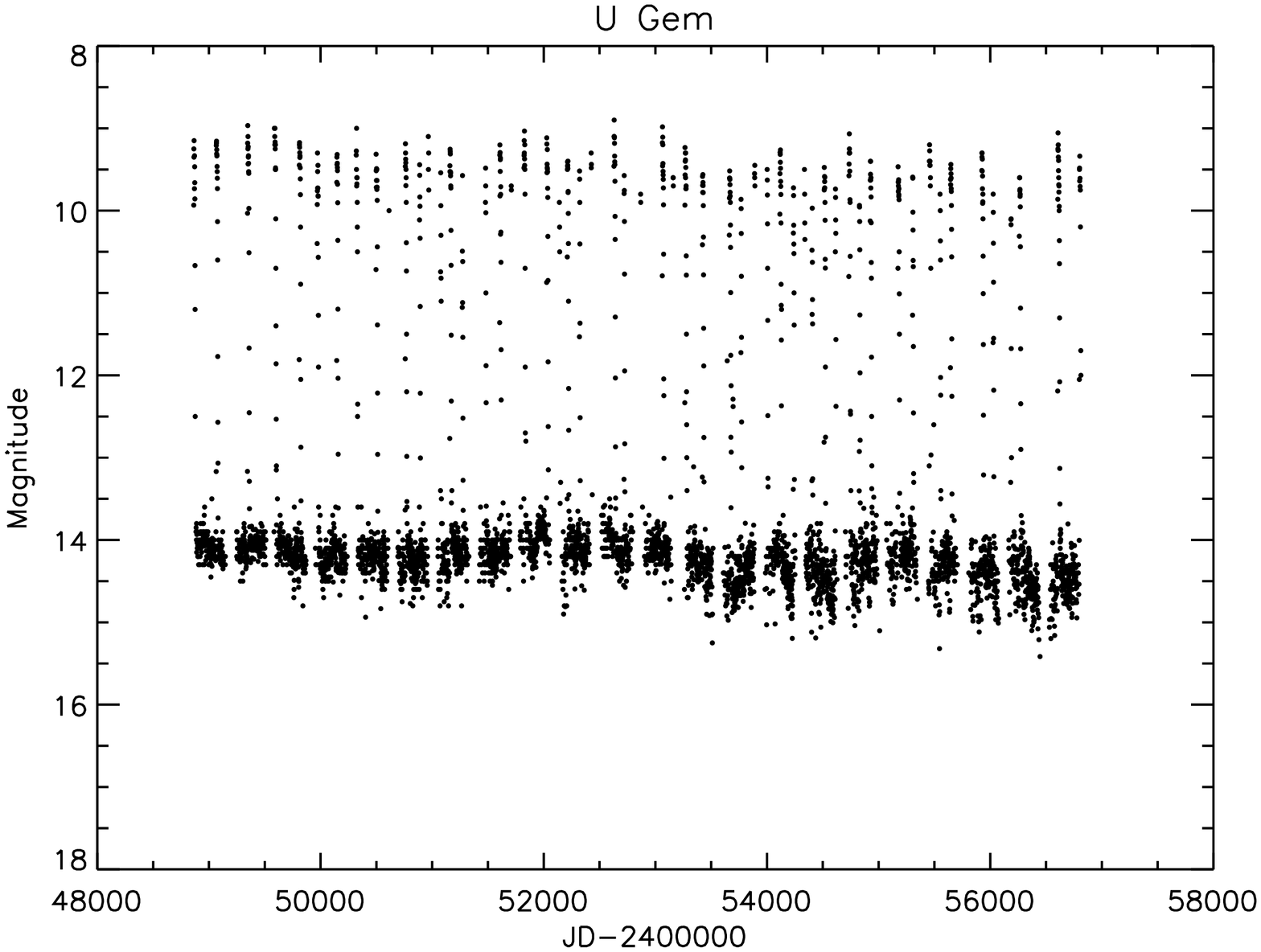}}
\subfigure{\includegraphics[width=0.25\textwidth,angle=90]{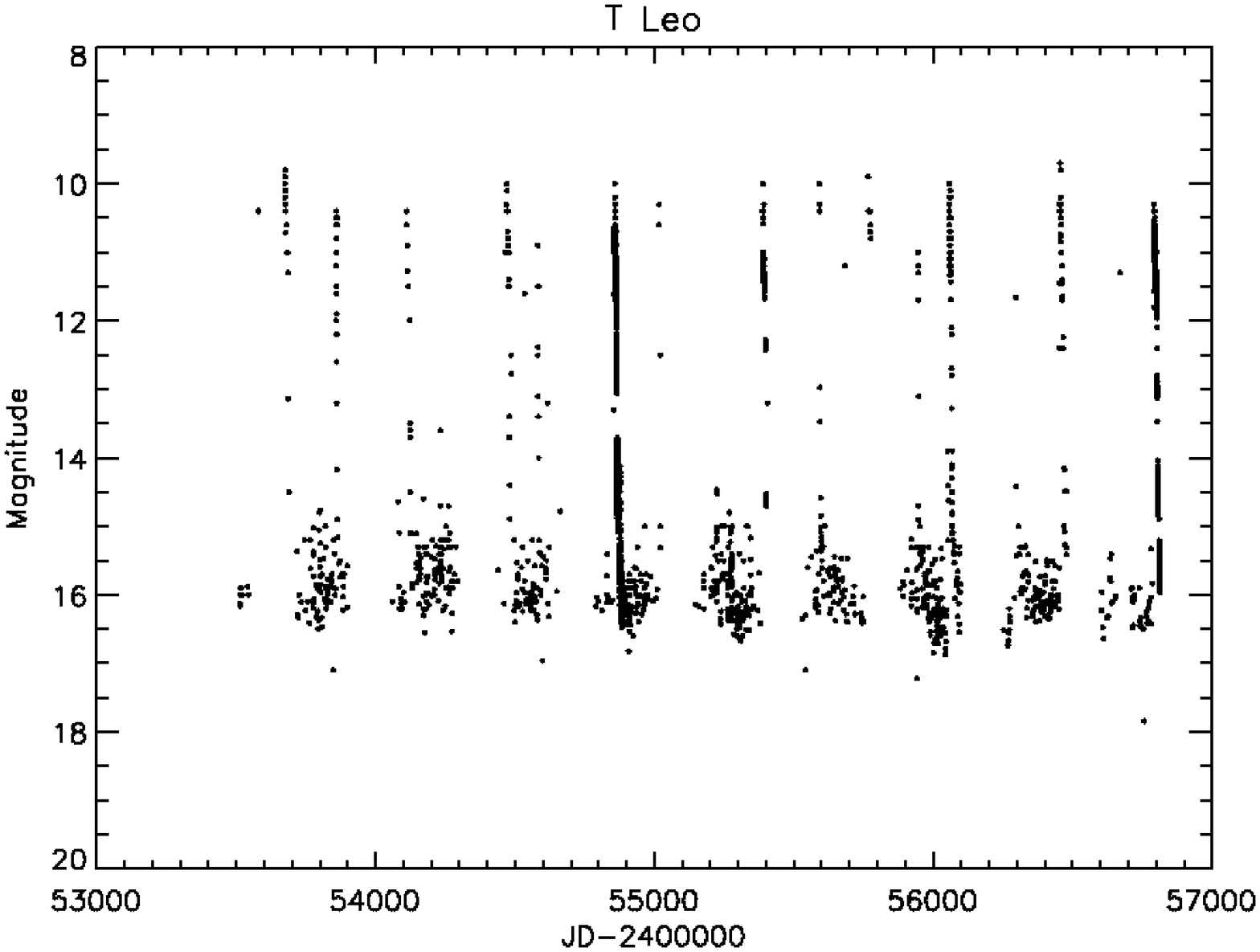}}\\
\subfigure{\includegraphics[width=0.25\textwidth,angle=90]{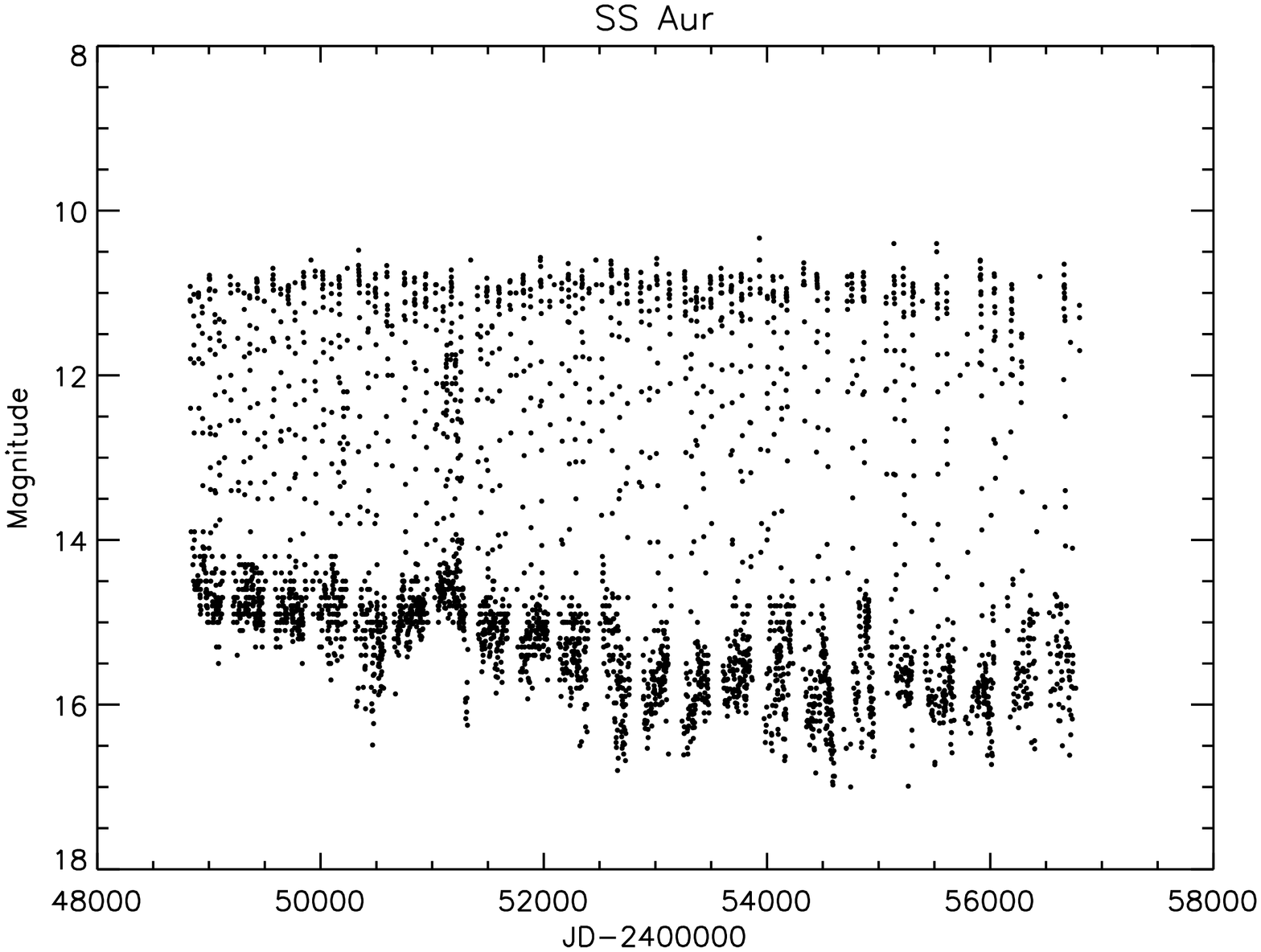}}
\subfigure{\includegraphics[width=0.25\textwidth,angle=90]{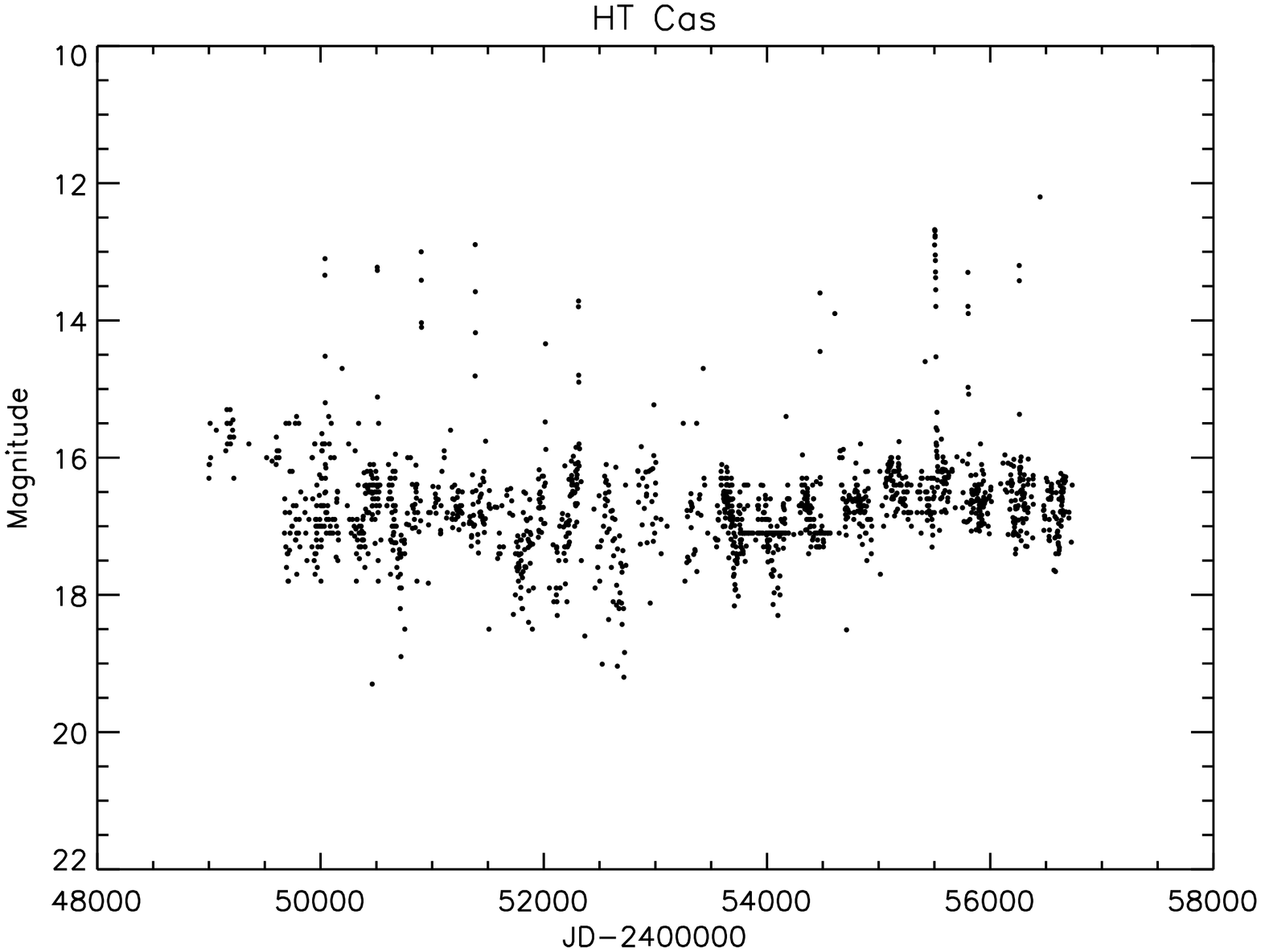}}
\subfigure{\includegraphics[width=0.25\textwidth,angle=90]{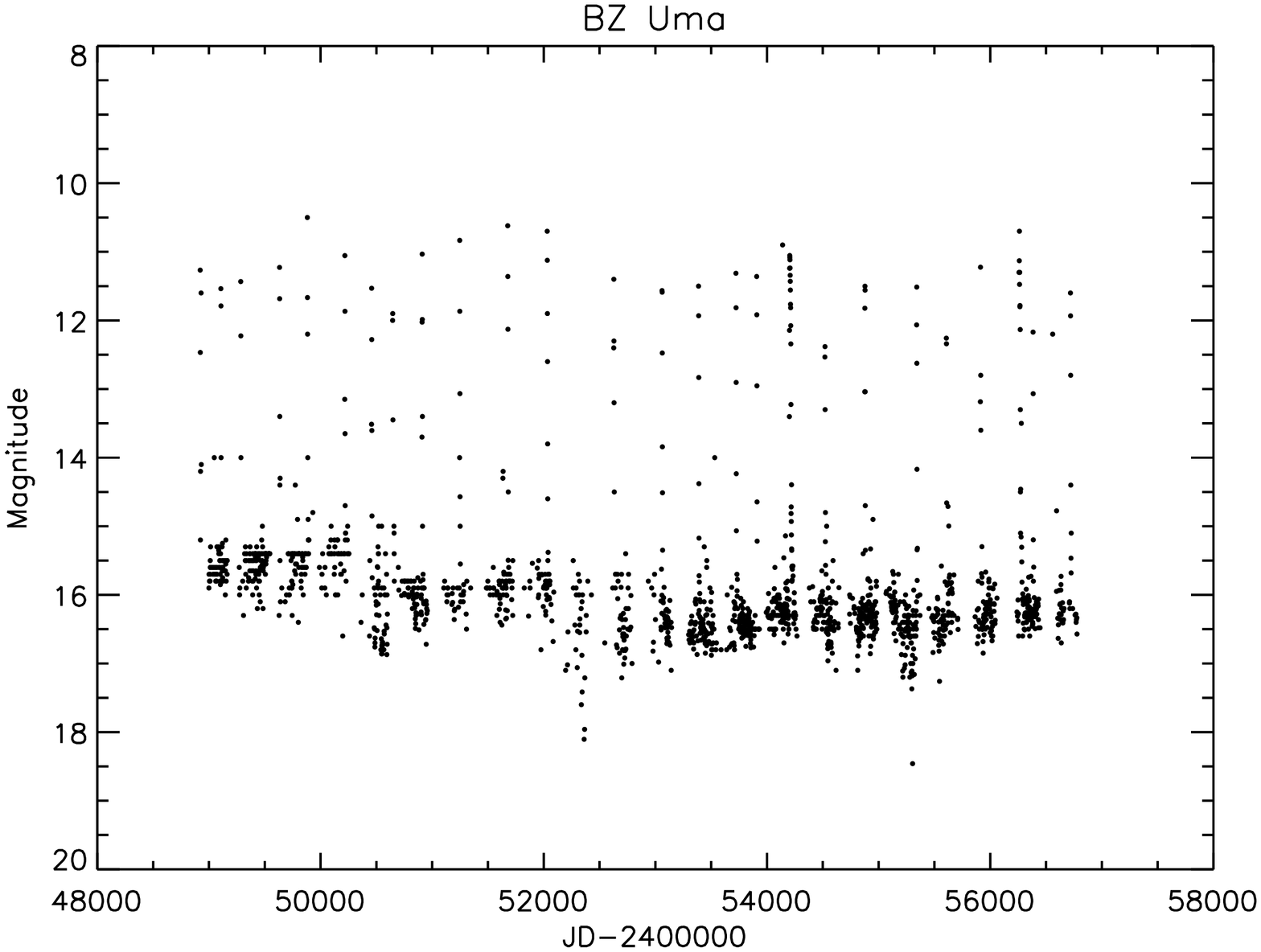}}
}
\caption{AAVSO lightcurves for DNe with parallax measurements within 200\,pc. Here we include only those sources with $L_{X}>3\times10^{30}\,\lum$ to avoid biases due to data incompletenesses at low $L_{X}$. Lightcurves for SS Cyg, U Gem, and Z Cam have been binned to 1 day averages for clarity.}
\label{fig:dnlcs}
\end{minipage}
\end{figure*}

Each of these CVs is well observed, with decades of publicly available data in multiple surveys. Where available, we use published values of duty cycles extant in the literature. For other objects we use AAVSO data in all available filters to measure the duty cycle of each CV in this sample, with results shown in Table \ref{tab:dntab}. The AAVSO data are generally accurate to within $0.2$ magnitudes for individual observations \citep{aavso} and is much better for many simultaneous observations. We define the duty cycle as the number of days during which the source is in outburst divided by the total baseline of observations. We define outburst as an event reaching at least 2 magnitudes greater than the quiescent magnitude beginning and ending when the object returns to $0.2$ magnitudes above the quiescent level prior to outburst. We do not include the standstills in Z Cam's high state as being in outburst; we consider only those times when Z Cam is in the low state and undergoing dwarf nova outbursts. We only include times at which the sources are visible and being observed consistently enough to be sure that no outbursts have been missed. Lightcurves of the data ranges used to measure duty cycle are presented in Figure \ref{fig:dnlcs}.

\begin{table*}
\caption{The measured duty cycles and extant X-ray luminosities for nearby dwarf novae. To fit a relation between duty cycle and X-ray luminosity, we use only those DN with $L_{X}>3\times10^{30}\,\lum$.}
\label{tab:dntab}
\begin{center}
\begin{tabular}{l c c c c}
System & $\log L_{X}$ {[2-10\,keV]} & Log Duty Cycle & $P_{orb}$ & References\textsuperscript{*} \\\relax
 & $(\lum)$ &  & (hr) & \\
\hline
BZ Uma & $31.17\pm0.23$ & $-1.39\pm0.08$  & 1.63 &  1 \\
HT Cas & $30.79\pm0.21$ & $-1.85\pm0.12$  & 1.77 &  1 \\
SS Aur & $30.982\pm0.095$ & $-0.61\pm0.07$ &  4.39 &  1 \\
SW Uma & $30.69\pm0.15$ & $-1.10\pm0.14$ &  1.36 & 1 \\
U Gem & $30.919\pm0.053$ & $-0.90\pm0.02$ &  4.25 & 1 \\
T Leo & $30.81\pm0.13$ & $-1.17\pm0.16$ &  1.42 & 1 \\
V893 Sco & $31.70\pm0.31$ & $-0.74\pm0.12$ &  1.82 & 1 \\
SS Cyg & $31.888\pm0.066$\textsuperscript{**}  & $-0.53\pm0.07$ &  6.603 & 2,3 \\
Z Cam & $31.79\pm0.33$ & $-0.38\pm0.05$ &  6.98 & 4 \\
\hline
GW Lib & $28.7\pm0.44$ & $-2.34\pm0.19$ &  1.28 & 1,5 \\
WZ Sge & $29.85\pm0.12$ & $-1.85\pm0.09$ &  1.36 & 1 \\
VY Aqr & $30.11\pm0.21$ & $-2.17\pm0.13$ &  1.51 & 1 \\
ASAS J0025 & $30.20\pm0.42$ & $<-1.92\pm0.15$ &  1.37 & 1,6 \\
\hline
\multicolumn{5}{l}{\textsuperscript{*} \footnotesize{References: (1) \citet{aavso} (2) \citet{SSCygDC} (3) \citet{Millerjones13}}} \\
\multicolumn{5}{l}{\footnotesize{(4) \citet{ZCamDC} (5) \citet{GWLib83} (6) \citet{ASAS}}} \\
\multicolumn{5}{l}{\textsuperscript{**} \footnotesize{X-ray luminosities come from \citet{Byckling10} in all cases but SS Cyg}} \\
\multicolumn{5}{l}{\footnotesize{where we use a revised distance based on radio parallax.}} \\
\end{tabular}
\end{center}
\end{table*}

\subsection{Fitting Methods}
\label{sec:fit}

Fitting a relation between the X-ray luminosities in the sample from 
\citet{Byckling10} and the corresponding duty cycles is complicated by the presence of intrinsic scatter around the relation that is not represented by the statistical errors. Methods for dealing with these difficulties along with their strengths and shortcomings are discussed in \citet{Tremaine02} and \citet{Kelly07}. To account for the intrinsic scatter of the data, we add a systematic uncertainty to all duty cycle measurements that is calculated such that $\chi^{2}/\nu=1$ for the initial fit by using the {\em MPFITEXY} routine \citep{Williams10} in the {\em MPFIT} package \citep{Markwardt09}. Without using this intrinsic scatter, $\chi^2/\nu=15$ on the final best fit, which is a decidedly poor fit. Because the errors in the measurement of X-ray luminosity are somewhat asymmetric, we symmetrize them in log space. The changes in the errors this causes are on the order of 10\%. We use a bootstrapping method to estimate the errors in the fit of the full data set, as the distribution of the bootstrapped fit results matches the error distribution of the best fit \citep{numericalrecipes}. For the bootstrapped fits without including an intrinsic scatter term, $\chi^2/\nu=11.8\pm8.4$, still well above the expected range of $\chi^2/\nu=1\pm\sqrt{2/\nu}$. In the bootstrapping analysis, the intrinsic scatter in the log of the duty cycle is on average $\sigma_{int}=0.31\pm0.12$, which gives $\chi^2/\nu=1.20$, which is within $\chi^2/\nu=1\pm\sqrt{2/\nu}$ for $\nu=7$. 

In order to estimate the errors in the relation, we fit the bootstrapped distribution of fits with a Gaussian and use the width of the Gaussian to estimate the error. The slope and intercept have been fit together in each iteration, with the intercept set to near the middle of the data to remove correlation between the intercept and the slope during the fitting.

\begin{figure}
\begin{center}
\includegraphics[width=0.35\textwidth,angle=90]{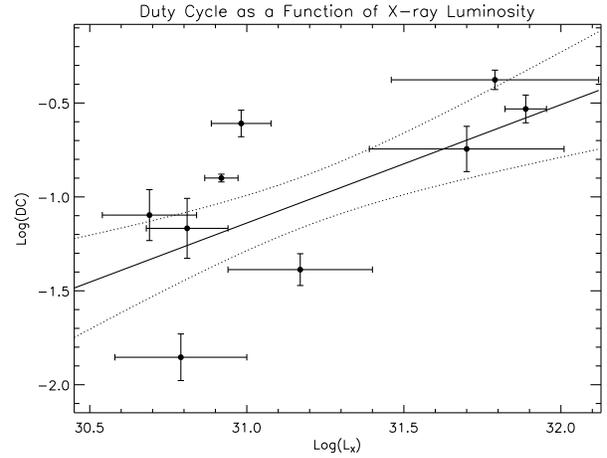}
\caption{
The best fit for Duty Cycle as a function of X-ray luminosity is plotted as a solid line. Dotted lines are the $1\sigma$ errors on this relation.}
\label{fig:dcvlx}
\end{center}
\end{figure}

The resulting best fit is:

$$\log DC=0.63\pm0.21\times(\log L_{X}(\lum)-31.3)-0.95(\pm0.1)$$

shown in Figure \ref{fig:dcvlx}. Fitting all of the DN in the \citet{Byckling10} sample gives a relation that is consistent with the one above, but which may be subject to selection effects against low duty cycle DN with $L_{X}<3\times10^{30}\,\lum$, and is given by:

$$\log{\rm DC}=0.65\pm0.08\times(\log L_{X}(\lum)-30.5)-1.50\pm0.14$$.

\begin{figure}
\begin{center}
\includegraphics[width=0.35\textwidth,angle=90]{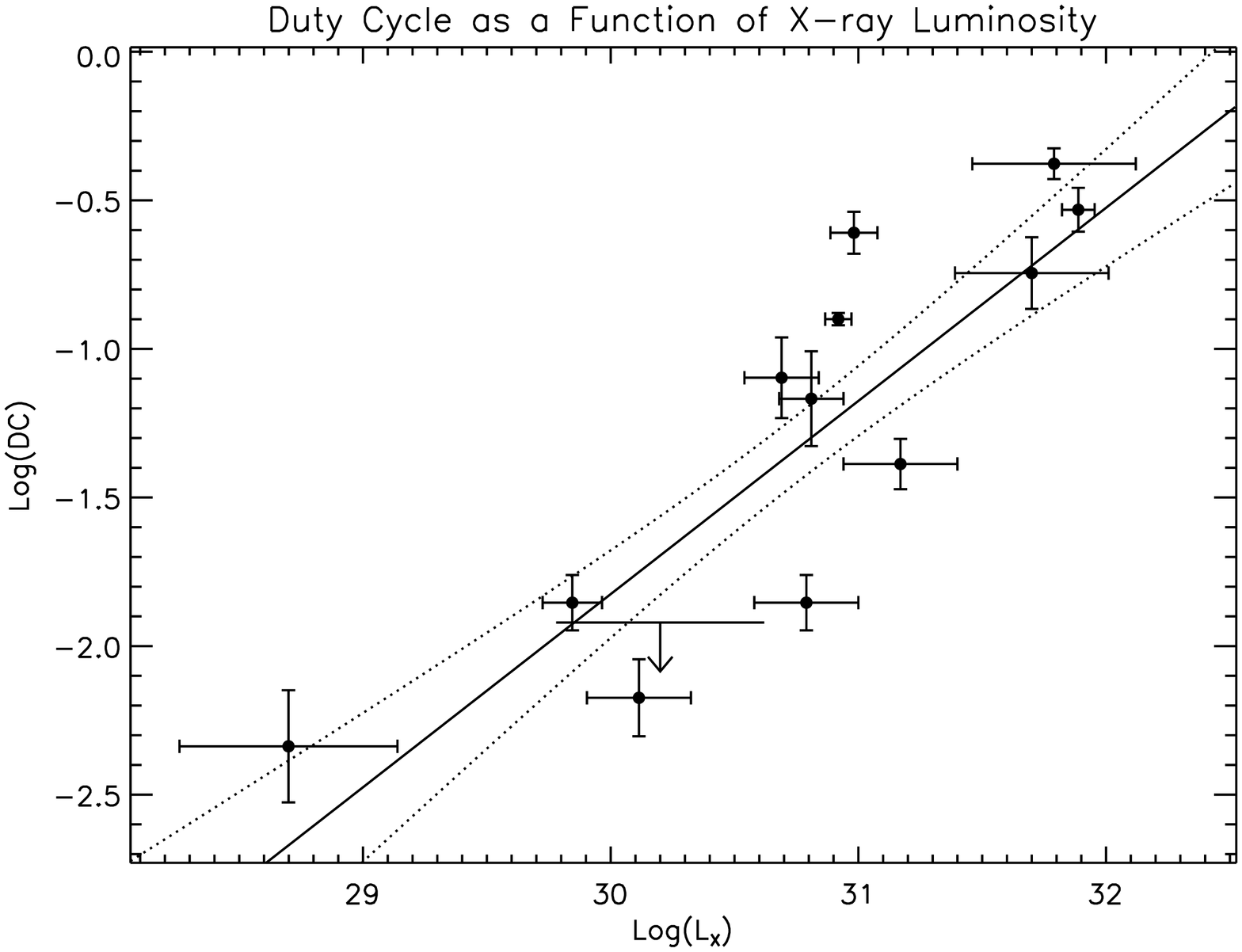}
\caption{The best fit (solid line) and $1\sigma$ errors (dotted lines) between the duty cycles and X-ray luminosities of the entire sample in \citet{Byckling10}. The fit to the entire sample agrees well with the fit to only the X-ray bright sample, but it is likely that there is a strong sampling bias against low duty cycle DN below $L_{X}=3\times 10^{30}\,\lum$.}
\label{fig:byck-all}
\end{center}
\end{figure}

This relation is unaffected by the inclusion or removal of ASAS J002511+1217.2 (ASAS J0025 henceforth), which has only 1 observed outburst, in the sample. The fits are shown in Figure \ref{fig:byck-all}.

\vspace{-2em}
\section{The Galactic Bulge Survey Sample of Dwarf Novae}

In the search for variability in candidate optical counterparts for the northern 3/4 of the GBS survey area, described in detail in \citet{Britt14}, 8 X-ray sources were observed to undergo a DN outburst. Of these 8, only the brightest (CX18) has so far been the target of both follow up spectroscopic observations and optical monitoring, where it was observed to undergo a subsequent DN outburst and to have an optical spectrum confirming that it is a non-magnetic CV \citep{Britt13}. 2 others, CX39 and CX87, have been targeted for spectroscopic follow-up which confirms their CV nature \citep{Torres14}. The remaining 5 DNe (CX: 81, 298, 426, 476, 860) have X-ray, infrared, and optical colors and flux ratios consistent, where available, with CVs undergoing DN outbursts in addition to the presence of a rise or decline of $>1$ magnitudes over a few days. Half of the observed DN occur in the brightest 100 X-ray sources out of the 1216 source survey population. The variability survey in \citet{Britt14} only examines stars coincident with X-ray positions in the GBS catalog; the data exist to study all objects in the field, but the full analysis is computationally intensive and beyond the scope of this paper. The X-ray detections for the GBS cover the energy range $[0.5-10]$\,keV, and include detections which have at least 3 photons.

The optical observations were made with the Mosaic-II instrument on the Blanco 4m telescope at CTIO in July, 2012. A description of the data reduction and photometry methods is given in \citet{Britt14}. The observations reach an optical depth of $r'=23$ and span an 8 day baseline, which is long enough to notice either the sharp rise or slower decay of a DN outburst.

In addition to the 8 objects classified as DN outbursts in \citet{Britt14}, there is one additional object, CX982, which could be a DN outburst with eclipses of the disk by the donor star during the outburst. It could also be some extreme coronal flaring, so we exclude it from our study here. If follow up observations reveal that it is indeed a DN, the space density estimated will be too low by $\sim12.5\%$. To account for the chance that this object should be included in the sample, and for the smaller but non-zero chance that 1 or more of our DNe in the sample are, in fact, not DNe, we include an additional systematic error of $12\%$ in our final estimation of space density.

\vspace{-2em}
\section{Estimating Space Density with Monte Carlo Simulations}

We can constrain the space density of CVs undergoing DN outbursts in the direction of the Galactic Bulge by modeling a CV population which follows the X-ray luminosity function of \citet{Pretorius12} and the duty cycle relation outlined above with various space densities, and comparing the resulting population with that seen by the GBS. We make several assumptions in this process.
\begin{itemize}
\item CVs trace stellar density in the disk, described by an exponential decay in both radius from the center and height above the plane of the disk. We have considered 3 possible scale lengths of the disk in our simulations, $2.2,3.0,3.8$\,kpc to account for the continuing uncertainty in the details of Galactic structure \citep{Juric08,Reyle09}. Changes to the scale length did not impact our results in a significant way. In addition, we have considered a range of scale heights of the disk. Because the scale height assumed had some impact on the results, we make the uncertainty of the true CV distribution a part of the simulation by adding a random component with a variance of 15 parsecs to a disc scale height centered on 120 parsecs \citep{Pretorius12}. We assumed a distance to the Galactic Center of $7620$\,pc \citep{Eisenhauer05}.
\item The peak absolute magnitude of a DN outburst is usually between 3 and 6, depending upon the orbital period and inclination \citep{Warner87,Patterson11}. An absolute outburst magnitude of 4.5 equates to an orbital period of 4.3 hours using the relation in \citet{Patterson11}. We use a Gaussian distribution of absolute magnitudes peaking at 4.5 with a variance of 1.5 magnitudes. To better estimate the effects of different distrubutions of absolute magnitudes, we reran the simulations with distributions peaking at 5.5 and 3.5 with the same variance. The changes in the detected population of DN outbursts are included in the errors in the final space density and specific frequency estimates. 

Because of extinction towards the Bulge, our optical survey would not recover an outburst with $M_{V}=4.5$ past a distance of $\approx5400$\,pc. Using a range of peak magnitudes as described above, we expect $\approx20\%$ of DNe to be further than this, quickly falling off with almost none as far as 8\,kpc. 
At $5-6$\,kpc in this longitude, the Bulge has started but contributes only $\sim15\%$ of the stellar population in this line of sight, though it rises quickly beyond this point. Using models of stellar density from the Bulge population compared to the disk population \citep{Stefano}, we find that the effect of including the Bulge in our models will change the number of detected DNe by $<10\%$, much smaller than uncertainties from small number statistics. With a better sample of DNe to work with, it would be worth using a more accurate model of the Milky Way. We have included a 10\% additional error in the estimate of specific frequency at the end of the process to address uncertainty from using a simplistic model of the Milky Way.  
\item The X-ray spectra of CVs are described by a thermal Bremsstrahlung model with a characteristic temperature of 15\,keV \citep{Mukai03}. Changing the characteristic temperature of the bremsstrahlung model by, e.g., a factor of 2 up or down makes little difference in the Chandra energy band, and thus has little effect on our calculations.
\item Extinction is roughly linear with distance (which is certainly not really the case since the lines of sight cross spiral arms, but this is a small source of error compared with the small number statistics), and is calculated assuming the extinction law from \citet{Clayton89}, the Vista Variables in Via Lactea reddening maps in \citet{Gonzalez12}, and the relation $N_{H}=0.58\times10^{22}$\,cm$^{-2}\times E(B-V)$ found in \citet{Bohlin78}, including the observational errors in the extinction maps which can be large in these regions.
\item X-ray detections of DNe in the GBS are made outside of outburst. This is clearly not accurate - for a ``worst case scenario'', we would expect for CVs in the GBS like SS Cyg, which is $\sim10\times$ brighter briefly at the start and end of a DN outburst and a high duty cycle (see Table \ref{tab:dntab}), that about $1/2$ would be detected in outburst due to the larger volume sampled, even though the brighter luminosity lasts only $\sim10\%$ of the duty cycle \citep{McGowan04}. Similarly, more systems like U Gem would be detected in outburst than in quiescence because the higher X-ray flux lasts throughout the duration of the outbursts. A survey that did not premise the detection of the outburst on an X-ray detection would escape this trouble, which could result in somewhat overestimating the specific frequency of CVs.
\end{itemize}
First, we throw a randomly generated population of CVs onto the region of the sky covered by the GBS and calculate the X-ray flux and optical magnitude of a DN outburst without correcting for extinction. The X-ray luminosities in \citet{Byckling10} are in a somewhat different energy band than the GBS data. We use WebPimms to correct for this difference using the assumptions about extinction and spectral shape listed above. We similarly correct for the different band used to calculate the X-ray luminosity function from \citet{Pretorius12}. If the flux values are above the GBS detection limits, we calculate the duty cycle based on the X-ray luminosity relation above, including the intrinsic scatter around this relation seen in the 9 DNe used for the fit. At every point that a relation is invoked, the uncertainty is included as a gaussian random number with a variance equal to the uncertainty in the relation. We then randomly place an observation at some point in the outburst cycle, and if the observation overlaps with a DN outburst, the DN is provisionally detected. At this point, we look up reddening values for the line of sight each provisionally detected DN is on and calculate extinction in both optical and X-ray, including the observational error in the extinction measures. If the CV remains above the flux limits of the GBS at this point, it is detected. 500 such simulations are run for different values of space density. We find that to generate a population of CVs undergoing DN outbursts consistent with those seen in the GBS, we require a space density of $\rho=5.6(\pm3.9)\times10^{-6}\,$pc$^{-3}$ in the solar neighborhood, scaled up with stellar density as we move closer to the Bulge. This is not a measurement of the local space density of high duty cycle CVs; it is a measurement of the density of high duty cycle CVs several kiloparsecs towards the Galactic Bulge, and is expressed in terms of the local space density under the assumption that the CV density through the Galaxy traces the stellar population. We write it in this way to allow a more direct comparison to published values of space density. An equivalent, but perhaps easier to vizualize, way of writing it is to use the local stellar mass density, $0.085\,M_{\odot}\,$pc$^{-3}$ \citep{McMillan11}, to state the average number of CVs per unit stellar mass in the parts of the Galactic disk being probed by the optical GBS survey, $6.6\pm4.7\times10^{-5}\,M_{\odot}^{-1}$. 

\begin{figure*}
\begin{center}
\includegraphics[width=0.35\textwidth,angle=90]{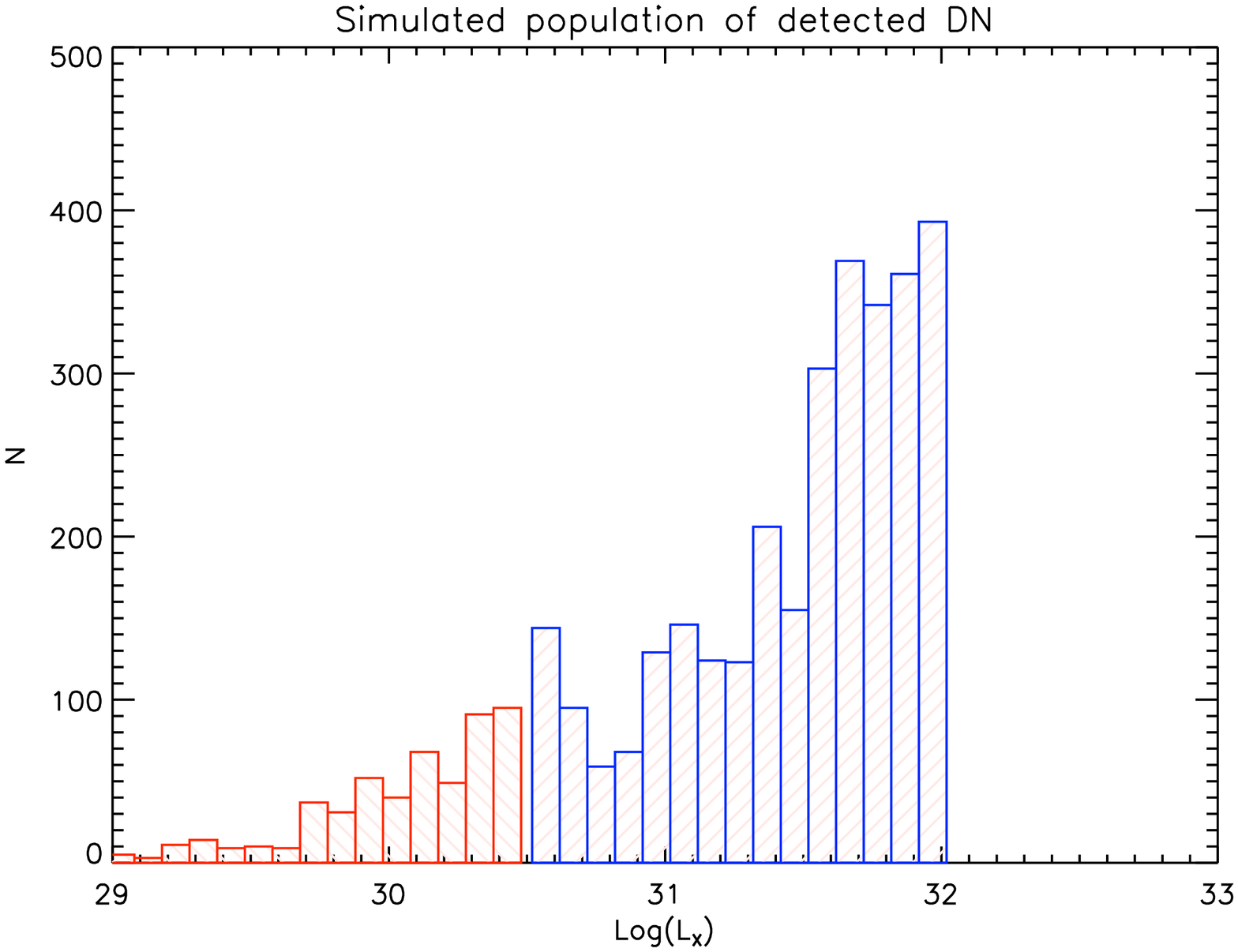}
\includegraphics[width=0.35\textwidth,angle=90]{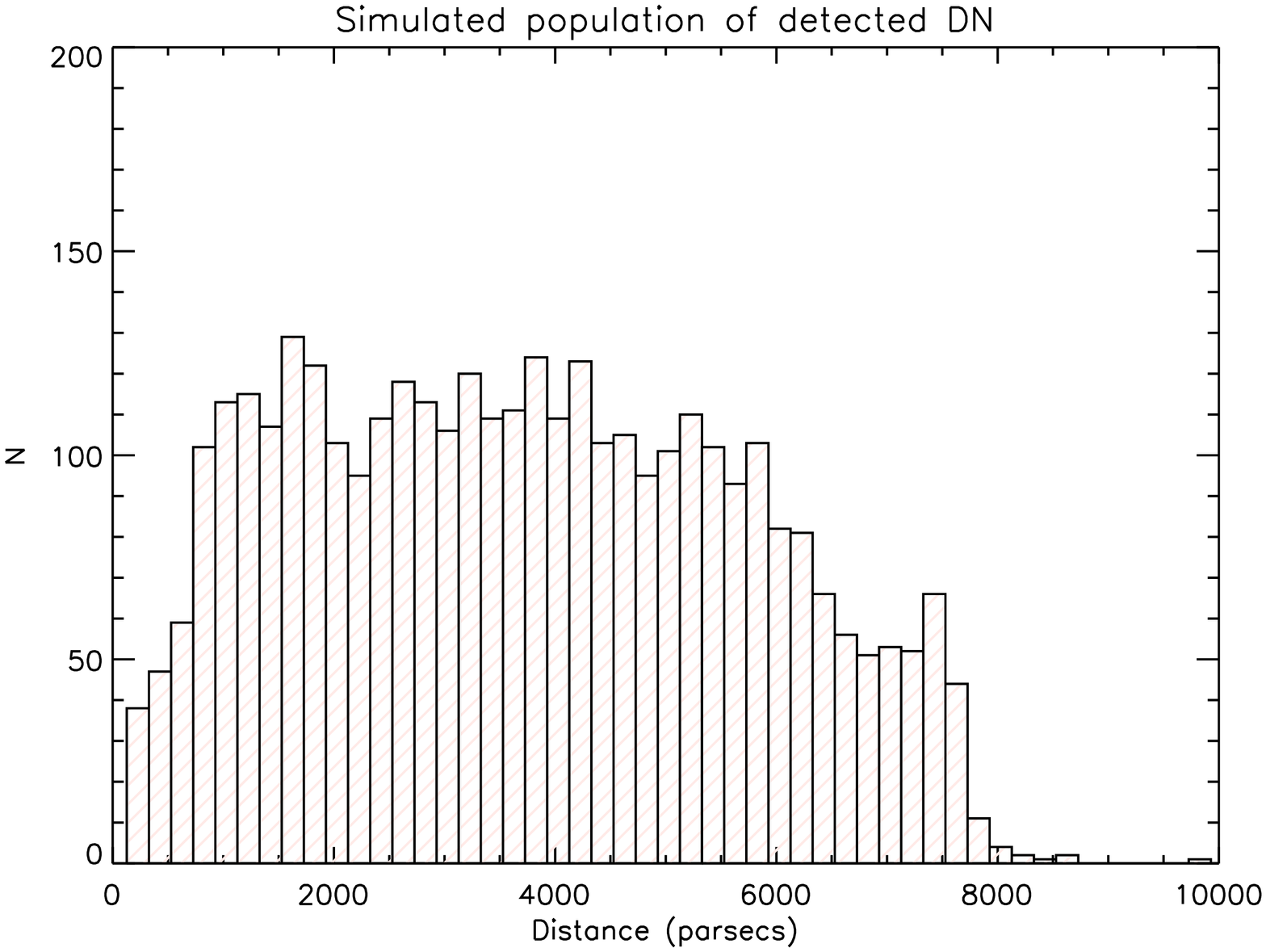}
\caption{{\it Left:} 
A histogram of the X-ray luminosities of the simulated population of DNe detected by the GBS using the relation between duty cycle and X-ray luminosity derived in Section \ref{sec:fit}. Systems below the break in the \citet{Byckling10} luminosity function are plotted in red with hatching at $-45^{\circ}$, while those above are in blue with hatching at $+45^{\circ}$. $84.8\%$ of systems are above the break. The GBS sample of DNe, therefore, likely is composed primarily of systems of comparable luminosity to those used to fit the relation between duty cycle and X-ray luminosity. Also and importantly, the sample of DNe detected in the GBS is relatively insensitive to the precise shape of the luminosity function, with space density of X-ray luminous CVs playing the major role. This is a result of the short baseline of GBS observations (only 8 days); longer surveys such as LSST will detect non-magnetic CVs with much lower duty cycles, offering a window into the distribution of the X-ray faint CV population. 
{\it Right:} The simulated distribution of distances of detected DNe. The optical depth of the survey combined with the bright X-ray luminosities expected from the very fact of the outburst detection in such a short baseline results in a distribution that falls off after $\sim6$\,kpc.
The primary limit to how far we can detect DN outbursts is the optical brightness of the outburst compared to our survey depth. 
}
\label{fig:synthlx}
\end{center}
\end{figure*}

This measurement is dominated by the high end of the X-ray luminosity function, as lower luminosity CVs undergo DN outbursts less frequently and are correspondingly less likely to be observed, in addition to the smaller volume reached. In particular, in our simulations of the DN population in the GBS which best match GBS observations only $15.2\%$ of detected DN have X-ray luminosities below the $L_{X}=3\times10^{30}\,\lum$ completeness limit of \citet{Byckling10}, as shown in Figure \ref{fig:synthlx}, which would correspond to roughly 1 of the 8 detected DN outbursts in the GBS.  As such, the exact shape of the assumed luminosity function at low to moderate luminosities has little effect on the resulting population, and the primary uncertainty comes ultimately from the small number of systems involved at both ends of the simulation. For example, using the luminosity function from \citet{Byckling10} instead does not have a significant impact on the resulting specific frequency. We also have very little in the way of a lever arm to constrain the relative size of the X-ray faint DN population to the X-ray bright DN population. The space density we measure is a factor of 2 higher than, but is still consistent with, the measurement of local long period DN in \citet{Pretorius12}, $2.1^{+3.5}_{-1.3}\times10^{-6}\,$pc$^{-3}$.

\vspace{-2em}
\section{Discussion}

Studies of the space density of CVs have concentrated on the local population, so that the changing density of the stellar environment as one moves outward radially in the disk is unimportant and generally not considered. In the GBS, however, we likely observe CVs undergoing DN outbursts out to $4-5$\,kpc. The radial and vertical exponential decay of the stellar density is therefore very important in our estimate of space density in this line of sight. GBS fields are also particularly close to the plane compared to other measures of space density of CVs which have tried to avoid extinction as much as possible by looking out of the plane of the Milky Way. The concept of a single ``space density'' for CVs is one that only makes sense in an extremely local environment where the density is relatively uniform, and is only a useful concept for the population at large in terms of a scale factor to the stellar density. 

While the non-uniform structure of the Galaxy is taken into account when calculating the space density of CVs, a more straightforward and ultimately more useful concept may be the specific frequency of CVs, which offers a more direct comparison to population models throughout different regions of the galaxy than the solar environment. The optical photometric follow-up to the GBS \citep{Britt14} reached a depth of $r'=23$. The optical depth of the survey combined with the bright X-ray luminosities expected from the very fact of the outburst detection in such a short baseline results in the distribution shown in the right panel of Figure \ref{fig:synthlx}. This work is therefore a probe of the density of non-magnetic CVs at distances from Earth much greater than any previous survey. Comparing our result to the local space density of long period CVs reached in \citet{Pretorius12}, there is some hint at a higher specific frequency of CVs as one moves in towards the center of the Galaxy; this is not yet a statistically significant effect, but we have a small number of DNe with which to work and make no attempt to factor in the distances of DNe in the GBS as a result. There is both a theoretical and observational basis to expect a change in specific frequency in different Galactic environments. \citet{Kundu02} show that an increase in the metallicity of Globular Clusters results in an increase in the specific frequency of Low Mass X-ray Binaries in those clusters. A possible reason for the increase is that donor stars with low metallicities do not develop convective zones at the same stellar mass, and so do not make contact until shorter orbital periods have been reached through other means of angular momentum loss such as magnetic braking \citep{Ivanova06}. We consider it likely, therefore, that the specific frequency of CVs should increase towards the Galactic Bulge as the metallicity increases, though these data are insufficient to show it. 

 This present work is severely limited by the small number of DNe available for the analysis in both the high luminosity Byckling sample and in the sample of DNe outbursts detected in the GBS in both optical and X-ray wavelengths. Though the DNe in the GBS sample are all detected in the X-ray, this method is primarily based on DN outburst detection, using the X-ray luminosity function and the connection between X-ray luminosity and Duty Cycle to connect to a predicted outburst rate. Our agreement with other space density estimates for DNe is a step forward for DN outburst detection methods, as other DN outburst surveys such as \citet{Cieslinski03} are in disagreement with other methods. Our result does not resolve any apparent tension between empirical and theoretical estimates of CV space density discussed in \autoref{intro}, as it agrees well with the majority of other empirical methods. 

LSST will offer a much better sampling of DNe systems. LSST will routinely go deeper in all parts of the sky, with frequent sampling and a long baseline that guarantees the detection of thousands of DN outbursts, including those with low duty cycles. Indeed, some large survey work revealing hundreds of new DNe outbursts is already forthcoming \citep[e.g.][]{Drake14}. Through this work, using large survey data to measure the frequency and brightness of outbursts for thousands of systems can yield an estimate of the relative proportion of CVs over all parts of the sky. Given the consistency of the optical brightness of DN outbursts \citep{Patterson11}, maps in 3 dimensions to distances of a few kpc could be made with no observations beyond what LSST is going to take anyway. A similar analysis with Pan-STARRS is complicated by the observing cadence on a similar or longer timescale than the duration of many DNe outbursts, many of which may only last a matter of days. 

The primary uncertainty of our estimate here comes from the small number of systems used, both to establish a quantitative relationship between duty cycle and X-ray luminosity and to match a synthetic population to our 8 DNe in the GBS optical campaign of \citet{Britt14}. The former is difficult to improve at present, as the number of X-ray luminous CVs for which highly accurate distances can be measured is small. However, the GAIA mission will undoubtedly improve this situation dramatically in several years' time, roughly concurrently with results from LSST \citep{Barstow14}. 

Neither the DN duty cycle nor the X-ray luminosity in quiescence are entirely reliable tracers of the secular mass accretion rate of the system (for which $\dot{M_{1}}=-\dot{M_{2}}$ in conservative mass transfer), as CVs at the same orbital period can exist in states of greatly differing mass transfer rates \citep[e.g. the nova-like below the period gap, BK Lyn;][]{Zellem09}. \citet{Buning04} suggest that this difference could be due to feedback cycles from irradiation driven mass transfer on the scale of $10^{4-5}$ years, but regardless of the mechanism, the disparate instantaneous mass transfer and accretion rates at equal orbital period is an observational fact. \citet{Patterson13} found that BK Lyn is transitioning to lower mass transfer rates, moving from a nova-like CV to a Z Cam-like state that alternates between a high and a low state, perhaps "calming down" after an historical nova in 101 A.D. quickly altered the equilibrium of the system.  Z Cam itself could be in a similar transitional period after a nova eruption \citep{Shara07}. If these periods of high mass transfer only last a few thousand years after a nova, then the majority of systems should have a strong relationship between mass transfer rate and orbital period. \citet{Patterson11} found that the recurrence time of DNe below the period gap was strongly a function of mass ratio, with $T_{rec}=318(\pm30)\,$days$\,(q/0.15)^{-2.63\pm0.17}$. The mass ratio also determines the Roche geometry $\frac{R_{L}}{a}$ and is strongly related to the WD temperature, itself a function of WD mass and accretion rate \citep[see Figure 10 of][]{Patterson11}. There is no theoretical prediction of the quiescent instantaneous accretion rate ($\dot{M_{1}}$) based on system parameters, though these observational signatures are suggestive. Our result may therefore be unsurprising to the community, but is the first quantitative relationship between the duty cycle and X-ray Luminosity, and so directly links the average mass transfer rate to the instantaneous accretion rate in quiescence. 
There is a strong selection effect on orbital period in our sample, as well, since long period systems have more massive donor stars \citep{Frank02} and there is a maximum WD mass. Above the period gap, the mass ratio $q$ must be higher than $q\approx0.25$, and will have relatively short recurrence times. Long period systems will therefore be significantly over-represented in a sample of X-ray luminous, high duty cycle DNe such as those in the GBS, though short-period systems are also sure to be present.

{\noindent\large\bf Acknowledgements}

COH is supported by NSERC, an Ingenuity New Faculty Award, and an Alexander von Humboldt Fellowship.
This research has made use of NASA's Astrophysics Data System Bibliographic 
Services and of SAOImage DS9, developed by Smithsonian Astrophysical 
Observatory.
We acknowledge with thanks the variable star observations from the AAVSO International Database contributed by observers worldwide and used in this research. We thank the referee, Koji Mukai, for his insightful comments improving the interpretation of this work.


\end{document}